\documentclass[11pt]{article}
\usepackage{epsfig,graphicx,amssymb,amsfonts}

\def \be {\begin{equation}}
\def \ee {\end{equation}}
\def \bea {\begin{eqnarray}}
\def \eea {\end{eqnarray}}

\def \dels {\partial\kern-.5em / \kern.5em}
\def \As {{A\kern-.5em / \kern.5em}}
\def \Ds {D\kern-.7em / \kern.5em}

\newcommand{\ena}{\end{eqnarray}}
\newcommand{\vs}[1]{\vspace{#1 mm}}

\def\bbox{{\,\lower0.9pt\vbox{\hrule \hbox{\vrule height 0.2 cm
\hskip 0.2 cm \vrule height 0.2 cm}\hrule}\,}}
\newcommand{\dsl}{\pa \kern-0.5em /}

\newcommand{\pa}{\partial}
\newcommand{\p}[1]{(\ref{#1})}

\def \K {{\tt I\kern-.25em K}}
\begin{document}
%\catcode`\@=11
%\catcode`\@=12
%\twocolumn[\hsize\textwidth\columnwidth\hsize\csname%
%@twocolumnfalse\endcsname

%%%%%%%%%%%%%%%%%%%%%%%%%%%%% LaTeX2e title page %%%%%%%%%%%%%%%
\begin{titlepage}

\begin{center}

\hfill\parbox{4cm}{
KU-TP 006 \\
{\normalsize\tt hep-th/0609043}
}

\vskip .5in

{\LARGE \bf Addendum to ``Hyperbolic Space Cosmologies''
%\\
%{}[JHEP {\bf 0310} (2003) 058]
}

\vskip 0.5in

{\large %
Chiang-Mei Chen$^a$\footnote{{\tt cmchen@phy.ncu.edu.tw}}, %
Pei-Ming Ho$^b$\footnote{{\tt pmho@phys.ntu.edu.tw}}, %
Ishwaree P. Neupane$^{c,d}$\footnote{{\tt
ishwaree.neupane@cern.ch}},
\\
Nobuyoshi Ohta$^e$\footnote{{\tt ohtan@phys.kindai.ac.jp}}
and John E.\ Wang$^{f}$\footnote{{\tt jwang@phys.cts.nthu.edu.tw}} }

\vskip 0.5in

${}^a$ {\it Department of Physics, National Central University,
Chungli 320, Taiwan}\\[3pt]
${}^b$ {\it Department of Physics, National Taiwan University,
Taipei 106, Taiwan}\\[3pt]
${}^c$ {\it Department of Physics and Astronomy, University of Canterbury,
Private Bag 4800, Christchurch 8020, New~Zealand}\\[3pt]
${}^d$ {\it Central Department of Physics, Tribhuvan University,
Kathmandu, Nepal}\\[3pt]
${}^e$ {\it Department of Physics, Kinki University,
Higashi-Osaka, Osaka 577-8502, Japan}\\[3pt]
${}^f$ {\it Physics Division, National Center for Theoretical Sciences,
Hsinchu, Taiwan}\\[3pt]

%{\normalsize June 2003}

\vs{15}
{\bf Abstract}
\end{center}
\vs{5}

In our earlier paper [JHEP 0310 (2003) 058], we considered higher dimensional
cosmological models with hyperbolic spaces. In particular the eternal
accelerating expansion was obtained by studying small perturbation around
the critical non-accelerated solution for $D > 10$. In this addendum,
we show that there is also such a solution in the critical case $D = 10$.

\end{titlepage}
\newpage

In our earlier paper~\cite{CHNOW}, we considered higher
dimensional cosmological models with hyperbolic spaces. In
particular the eternal accelerating expansion was analyzed by
studying small perturbations around the critical non-accelerated
solution. We concluded that eternal acceleration is possible
for $D > 10$. In this note we want to show that there is also
such a solution in the case $D = 10$ as well, which was overlooked
in~\cite{CHNOW}. The existence of such a solution was recently
argued in refs.~\cite{AH, ST}.

For this critical dimension, our metric is
\begin{equation}
ds^2 = \mathrm{e}^{-6\phi} \left( -dt^2 + a^2 ds_{H_3}^2 \right) +
\mathrm{e}^{2\phi} ds_{H_6}^2,
\end{equation}
where $ds_{H_n}^2$ denotes the metric of an $n$-dimensional
hyperbolic space.
The critical solution is
\begin{equation}
a = 2 t, \qquad \phi = \frac1{2\sqrt6} \psi + \frac18 \ln 30,
\label{cri1}
\end{equation}
where
\begin{equation}
\psi = \frac1{c} \ln \left(\frac{t^2}{3}\right), \qquad c = 2
\sqrt{\frac23}.
\label{cri2}
\end{equation}

Previously in our paper \cite{CHNOW} we obtained accelerating
solutions by perturbing the above solution as
\begin{equation}
a = a_0 + a_1, \qquad \psi = \psi_0 + \psi_1,
\end{equation}
where $a_0$ and $\psi_0$ are the critical solutions in \p{cri1} and \p{cri2}.
The perturbative parts were written as
\begin{equation}
a_1 = A, \qquad \psi_1 = \frac{B}{t}
\end{equation}
where $a_1$ and $\psi_1$ were chosen to be polynomials which were then
fixed using the equations of motion. For perturbation theory to be
valid, it was necessary to assume that $a_1 \ll a_0$ and $\psi_1
\ll \psi_0$. Our ansatz led to solutions with accelerated expansion
behavior for generic $D > 10$. For the case $D=10$ the analysis resulted
in constant $A$ and $B$, and therefore there was no eternal acceleration.
However, we now re-examine this case and perform more detailed
analysis for the case $D=10$ allowing the possibility that $A$
and $B$ depend on time in search of accelerated expansion
behavior.

Our approach is to re-examine the equations of motion for
$D=10$ and directly look for solutions starting from the above
ansatz with arbitrary functions $A$ and $B$. The equations of
motion for the first order term (given as eqs.~(5.11)-(5.13)
in~\cite{CHNOW}) reduce in this particular case to
\begin{eqnarray}
\dot A - \frac{3A}{4t} - \frac{c}{4} \dot B + \frac{3c}{4}
\frac{B}{t} &=& 0,
\label{1} \\
\ddot B + \frac{\dot B}{t} + \frac{3B}{t^2} + \frac{3}{c} \left(
\frac{\dot A}{t} -\frac{A}{t^2} \right) &=& 0,
\label{2} \\
\ddot A + \frac{c}{t} \dot B &=& 0. \label{3}
\end{eqnarray}
To obtain a solution we first eliminate $A$ from eqs.~(\ref{1})
and (\ref{2}) and get as a result
\begin{equation}
\dot A - c t \ddot B - 2 c \dot B = 0, \label{4}
\end{equation}
where the dot denotes the derivative with respect to time. If we
differentiate this result once and use eq.~(\ref{3}) to eliminate
the variable $A$, we obtain the equation for $B$:
\newcommand{\dddot}[1]{\stackrel{...}{#1}}
\begin{equation}
t^2 \dddot B + 3 t \ddot B + \dot B = 0,
\end{equation}
whose solution is
\begin{equation}
B = c_1 (\ln t)^2 + c_2 \ln t + c_3.
\end{equation}
Finally we obtain the solution for the variable $A$ by
substituting the solution for $B$ into eq.~(\ref{4})
\begin{equation}
A = c \left[ c_1 (\ln t)^2 + (2 c_1 + c_2) \ln t + c_4 \right].
\end{equation}
Consistency of the equations tells us that
\begin{equation}
c_4 = \frac{8}{3} c_1 + c_2 + c_3.
\end{equation}

The expansion and acceleration of the cosmic evolution are
\begin{equation}
\dot a = 2 + c t^{-1} (2 c_1 \ln t + 2 c_1 + c_2), \quad \ddot a =
- c t^{-2} ( 2 c_1 \ln t + c_2 ).
\end{equation}
The acceleration occurs for all
\begin{equation}
2 c_1 \ln t + c_2 < 0, \qquad \mathrm{or} \qquad t >
\mathrm{e}^{-c_2/2c_1},
\label{cond}
\end{equation}
if we choose $c_1 < 0$. The latter condition is necessary for the accelerated
expansion to continue in the future direction.
Since the expanding condition, which is also the condition for the
perturbation theory to be valid,
\begin{equation}
t > - \frac{c}2 (2 c_1 \ln t + 2 c_1 + c_2),
\end{equation}
can always be satisfied for sufficiently large $t$, the accelerated
expanding phase always occurs after certain time of  evolution

Although $a_1$ approaches infinity as $t \rightarrow \infty$, the
first order perturbative treatment above is still reliable for
large $t$ because $\ln t / t \ll 1$ and we still have $a_1 \ll
a_0$ and $\psi_1 \ll \psi_0$ for all $t$. Thus hyperbolic
compactification can give eternal acceleration for $D\geq 10$
instead of just $D>10$.
% Indeed there exist various other possibilities; see figures
%\ref{hyper1}-\ref{hyper6}.

As a final comment, our solution is related to eq.~(3.19) in
\cite{AH} via the time redefinition
\begin{equation}
t = \left(\frac58\right)^{3/2} \frac14 \left[ \tau^4 - 18 c_0 (\ln
\tau)^2 - 36 d_0 \ln \tau \right],
\end{equation}
where $\tau$ is the time coordinate and $c_0, d_0$ are parameters
of the result in \cite{AH}. Here the factor $5/8$ comes from the
different conventions for hyperbolic space by a constant scale
factor. Our seemingly extra parameter $c_3$ is a redundancy and
%can be absorbed by coordinate redefinition into scaling the
%value of the origin of time $t$.
%Moreover, our solution has one more parameter, i.e. $c_3$
%which is absent in \cite{AH}. It
can be absorbed by the coordinate transformation of scaling $t$.

\section*{Acknowledgements}

IPN would like to thank the CERN Theory Division for their kind
hospitality during the course of this work. This work was carried
out while NO was visiting the Institute of Theoretical Physics,
Chinese Academy of Sciences and Centre for Mathematical Sciences,
Cambridge University, and he would like to thank R.-G. Cai and
P.K. Townsend for valuable discussions and kind hospitality. The
work of CMC, PMH and JEW is supported in part by the National
Science Council, the Center for Theoretical Physics at National
Taiwan University, the National Center for Theoretical Sciences,
Taiwan, R.O.C. That of IPN is supported in part by the Foundation
for Research, Science and Technology (New Zealand) under Research
Grant No. E5229. That of NO is supported in part by Grant-in-Aid
for Scientific Research Fund of the JSPS No. 16540250.  That of
JEW is supported in part by the Academic Center for Integrated
Sciences at Niagara University and the New York State Academic
Research and Technology Gen``NY''sis Grant.

%%%%%%%%%%%%%%%%%%%%%%%%%%%%%%%%%%

\end{document}